\documentclass[aps,pre,preprint,showpacs,superscriptaddress]{revtex4}

\def \beq{\begin{equation}}         \def \eeq{\end{equation}}
\def \beqa{\begin{eqnarray}}        \def \eeqa{\end{eqnarray}}
\def \bea{\begin{array}}        \def \eea{\end{array}}

\def\bioz#1#2#3{{Biochem. Z. }{\bf #1}, #2 (#3)}

\def\jpc#1#2#3{{J. Phys. Chem. B. }{\bf #1}, #2 (#3)}

\def\jcp#1#2#3{{J. Chem. Phys. }{\bf #1}, #2 (#3)}

\def\natc#1#2#3{{Nature chem. Biol. }{\bf #1}, #2 (#3)}
\def\pnas#1#2#3{{Proc. Natl. Acad. Sci. USA }{\bf #1}, #2 (#3)}

\def\sci#1#2#3{{Science }{\bf #1}, #2 (#3)}

\usepackage{amsmath}
\usepackage{epsfig}
\begin{document}

\title{Single Molecule Michaelis-Menten Equation beyond Quasi-Static Disorder}
\author{Xiaochuan Xue}
\affiliation{Center for Advanced Study, Tsinghua University,
Beijing, 100084, China}
\author{Fei Liu}
\email[Email address:]{liufei@tsinghua.edu.cn} \affiliation{Center
for Advanced Study, Tsinghua University, Beijing, 100084, China}
\author{Zhong-can Ou-Yang}
\affiliation{Center for Advanced Study, Tsinghua University,
Beijing, 100084, China} \affiliation{Institute of Theoretical
Physics, The Chinese Academy of Sciences, P.O.Box 2735 Beijing
100080, China}

\date{\today}

\begin{abstract}
The classic Michaelis-Menten equation describes the catalytic
activities for ensembles of enzyme molecules very well. But recent
single-molecule experiment showed that the waiting time
distribution and other properties of single enzyme molecule are
not consistent with the prediction based on the viewpoint of
ensemble. It has been contributed to the slow inner conformational
changes of single enzyme in the catalytic processes. In this work
we study the general dynamics of single enzyme in the presence of
dynamic disorder. We find that at two limiting cases, the slow
reaction and nondiffusion limits, Michaelis-Menten equation
exactly holds although the waiting time distribution has a
multiexponential decay behaviors in the nondiffusion limit.
Particularly, the classic Michaelis-Menten equation still is an
excellent approximation other than the two limits.
\end{abstract}
\pacs{87.14.Ee, 82.37.-j, 05.40.-a, 87.15.Aa } \maketitle

The Michaelis-Menten (MM) mechanism~\cite{Michaelis} is widely
used to understand the catalytic activities of various enzymes.
According to this mechanism, a substrate S binds reversibly with
an enzyme E to form a complex ES. ES then undergoes unimolecular
decomposition to form a product P, and E is regenerated for the
next cycle.
\begin{equation}
\label{MMmechanism} \textrm{E+S}\ \
\raisebox{-2.0ex}{$\stackrel{\stackrel{\textrm{\small $k_1$}}
{\textrm{\Large $\rightleftharpoons$}}}{\textrm{\small
$k_{-1}$}}$}\ \ \textrm{ES}\xrightarrow{{k_2}} {\rm E}^0+{\rm P},\
\ \ {\rm E}^0\xrightarrow{{k}} {\rm E}
\end{equation}
The rate of product formation $v$ on substrate concentration $[S]$
can be characterized by the MM equation~\cite{Michaelis}
\begin{eqnarray}
\label{MMequ} &&v=\frac{v_{\rm max}[S]}{[S]+K_{\rm M}}
\end{eqnarray}
where $v_{\rm max}=k_2[E]_{\rm T}$ is the maximum generation
velocity, $[E]_T=[E]+[ES]$ is the total enzyme concentration, and
$K_{\rm M}=(k_{-1}+k_P)/k_1$ is the Michaelis constant. Although
the MM mechanism and equation have been found for almost a hundred
years, they are still widely accepted and remain pillars of
enzymology.

Even if the classic MM equation achieves considerable success,
there are still many intriguing problems about the equation
waiting to be answered. Particularly, the recent single-molecule
fluorescence studies~\cite{Lu,Zhuang,Oijen,Yang,Flomenbom} found
that catalytic rates of many enzymes are fluctuating with time due
to conformational fluctuations. A natural question hence is why MM
equation works well despite the broad distributions and dynamic
fluctuations of single-molecule enzymatic rates. Recently Xie {\it
et al.} tried to address this issue from view of points of
single-molecule experiment~\cite{English} and theory~\cite{Kou}.
In addition that the reciprocal of the first moment of $f(t)$,
$\langle t \rangle^{-1}=v/[E]_{\rm T}$ follows MM equation well at
any substrate concentration, the most remarkable discovery of
their experiment is that the waiting time distributions $f(t)$
exhibit highly stretched multiexponential decays at high substrate
concentration and monoexponential decays at low substrate
concentration~\cite{English}. Xie {\it et al.}~\cite{Kou}
attributed the nonexponential decay of $f(t)$ to dynamic disorder
of the rate constants of the reactions in Eq.~(\ref{MMmechanism})
caused by transitions among different enzyme conformations. They
theoretically proved that the classic MM equation still holds at
the single molecular level when the transition rates among the
$ES$ conformations are slower than the catalytic rate $k_2$ (the
quasi-static condition), even if $f(t)$ is no longer
monoexponential decays at high substrate concentrations. Therefore
one of following issues is whether we can still derive the MM
equation beyond the quasi-static disorder. Xie {\it et
al.}~\cite{Kou} indeed attempted to give an answer about it. But
their effort ended in the two-state model for the algebraically
complex in the multistate model. In this work, we propose that the
classic MM equation holds under broader disorder conditions.
Different from the discrete state model of Xie {\it et al.}, a
continuum diffusion-reaction model is used~\cite{AgmonHopfield}.

The conformational probability density for each enzyme state,
$P_{\rm I}(x,t)$, in Eq.~(\ref{MMmechanism}) can be obtained by
three coupled diffusion-reaction equations with the potential
[$V_{\rm I}(x) $] and the reaction terms [$k_i(x)$]
\begin{eqnarray}
\label{diffusionreactonequations}&& \frac{\partial}{\partial
t}P_{\rm E}(x,t)=\left[{\cal L}_{\rm E}-k_{1\rm S}(x)\right]P_{\rm E}+k_{-1}(x)P_{\rm ES}\nonumber\\
&&\frac{\partial}{\partial t}P_{\rm ES}(x,t)=\left[{\cal L}_{\rm
ES}-k_3(x)\right]P_{\rm ES}+
k_{1\rm S}(x)P_{\rm E}\\
&&\frac{\partial}{\partial t}P_{\rm E^0}(x,t)={\cal L}_{\rm
E^0}P_{\rm E^0}+k_2(x)P_{\rm ES}\nonumber
\end{eqnarray}
where
\begin{eqnarray}
\label{definitionFKoperator}{\cal L}_{\rm I}=D_{\rm
I}\frac{\partial}{\partial x}\exp[-\beta V_{\rm
I}(x)]\frac{\partial}{\partial x}\exp[\beta V_{\rm I}(x)],
\end{eqnarray}
and I$=$E, ES or E$^0$, and $k_3(x)=k_{-1}(x)+k_2(x)$ and $k_{1
S}(x)=k_1(x)[S]$ are defined for convenience. The diffusion
coefficient $D_{\rm I}$ determines the rate of the conformational
transition on the state $i$. The initial conditions are $P_{\rm
ES}(x,0)=0$, $P_{\rm E^0}(x,0)=0$, and $P_{\rm E}(x,0)$ is the
thermal equilibrium distribution with the potential $V_{\rm
E}(x)$. In single molecule turnover experiment, the observation is
the probability density of the waiting time for an enzymatic
reaction $f(t)$, which is defined
\begin{eqnarray}
\label{waitingtimedistributionLP} f(t)=\int k_2(x)P_{\rm ES}(x)dx.
\end{eqnarray}

We first study the solutions of
Eq.~(\ref{diffusionreactonequations}) in two limiting cases: the
slow reaction and the nondiffusion limits. In addition that the
coupled diffusion-reaction equations have exact analytical
solutions under these limits, this study would be useful in
understanding the general solutions of
Eq.~(\ref{diffusionreactonequations}). Particularly, we will show
that quasi-static condition proposed by Xie {\it et al.} is just
one of case of the latter limit. \\
{\bf  The slow reaction limit} In this limit the processes of
reactions is very slowly compared to processes of the enzyme
conformational diffusion. Therefore the thermal distributions are
always maintained during the courses of reactions. The solution to
the diffusion-reaction equations then can be written as
\begin{eqnarray}
P_{\rm I}(x,t)=P^{\rm eq}_{\rm I}(x)\rho_{\rm I}(t).
\end{eqnarray}
where $P^{\rm eq}_{\rm I}(x)\propto \exp[-\beta V_{\rm I}(x)]$,
and $\beta^{-1}=k_{\rm B}T$, $k_{\rm B}$ is the Boltzmann's
constant, and $T$ is absolute temperature. Substituting them into
Eq.~(\ref{finalwaitingtimedistributionLP}) and considering that
\begin{eqnarray}
{\cal L}_{\rm I}P^{\rm eq}_{\rm I}(x)=0,
\end{eqnarray}
after simple calculations we get
\begin{eqnarray}
\label{slowreactionlimit} f(t)&=&\rho_{\rm ES}(t)\int k_2(x)P^{\rm
eq}_{\rm ES}(x)dx\\
&=&\frac{k^{1S}_{{\rm E}_{\rm eq}}k^2_{{\rm ES}_{\rm
eq}}}{2A_e}\left[e^{(B_{\rm eq}+A_{\rm eq})t}-e^{(B_{\rm
eq}-A_{\rm eq})t}\right],\nonumber
\end{eqnarray}
where $A_{\rm eq}=\left[(k^3_{{\rm ES}_{\rm eq}}+k^{1S}_{{\rm
E}_{\rm eq}} )^2/4-k^{1S}_{{\rm E}_{\rm eq}}k^2_{{\rm ES}_{\rm
eq}}\right]^{1/2}$ and $B_{\rm eq}=-(k^3_{{\rm ES}_{\rm
eq}}+k^{1S}_{{\rm E}_{\rm eq}})/2$. Hence the reciprocal of the
mean waiting time is
\begin{eqnarray}
\label{slowreactionlimitMM} \frac{1}{\langle
t\rangle}=\frac{k^2_{{\rm ES}_{\rm eq}}[S]}{[S]+M_{\rm eq}},
\end{eqnarray}
where $M_{\rm eq}=(k^{-1}_{{\rm ES}_{\rm eq}}+k^2_{{\rm ES}_{\rm
eq}})/k^1_{{\rm E}_{\rm eq}}$. We can see that in this rapid
diffusion limit, Eq.~(\ref{slowreactionlimitMM}) is almost the
same as the single molecule MM equation in the absence of dynamic
disorder~\cite{Kou} except that the rate constants now are the
mean values on their inner conformational coordinate.\\
{\bf The nondiffusion limit ($k_i^{-1}\ll\beta D_{\rm I}$)} In
this limit the reactions in Eq.~(\ref{MMmechanism}) proceed so
rapidly at the initial values of the slow coordinate $x$ that the
distribution of $x$ is not restored by diffusion in the course of
reactions. Then the diffusion terms in the diffusion reaction
equations are neglected or $D_{\rm I}\approx0$. The following
calculations are simple and we immediately have
\begin{eqnarray}
\label{nondiffusionlimitdistribution} f(t)=\int P^{\rm
eq}_E(x)\frac{k_{1S}(x)k_2(x)}{2A(x)}\left\{e^{[B(x)+A(x)]t}-e^{[B(x)-A(x)]t}\right
\}dx
\end{eqnarray}
and
\begin{eqnarray}
\label{nondiffusionlimitMMeq}
\frac{1}{\langle
t\rangle}=\frac{\kappa_{\rm nd}[S]}{[S]+M_{\rm nd}}
\end{eqnarray}
where
\begin{eqnarray}
&&\kappa_{\rm nd}^{-1}= \int dxP^{\rm eq}_{\rm
E}(x)/k_2(x)dx, \nonumber \\
&&M_{\rm nd}=\kappa_{\rm nd}\int P^{\rm eq}_{\rm
E}(x)k_3(x)/\left[k_1(x)k_2(x)\right]dx, \nonumber
\end{eqnarray}
where
$A(x)=\left[(k_{1S}(x)+k_3(x))^2/4-k_{1S}(x)k_2(x)\right]^{1/2}$
and $B(x)=-(k_3+k_{1S})/2$.

We note that the expressions of the waiting distribution and the
mean waiting time in the latter limit is very similar with the
main conclusion [Eq. (31)] obtained by Xie {\it et
al.}~\cite{Kou}. It is not unexpected because the quasi-static
condition used by Xie {\it et al.} is included in our nondiffusion
limit. But two new points are revealed in the present work. One is
that, in addition to $k_2$, the other rates may also be allowed to
fluctuating in time. The other and more interesting finding is
that the unknown steady-state weight function $w(k_2)$ introduced
\emph{in prior} by Xie {\it et al.} has a direct physical
interpretation. To better understand the similarity between our
calculation with them, we rewrite
Eq.~(\ref{nondiffusionlimitdistribution}) by viewing $k_2$ as
variable instead of $x$, and make use of the experimental
observations~\cite{English} that both $k_1(x)$ and $k_{-1}(x)$ are
independent of the conformational coordinate (the ``wide reaction
window" limit termed by Sumi and Marcus~\cite{Sumi}), then we
obtain the same expression as that Xie \emph{et al.} solved by
very complicated algebra operations.
\begin{eqnarray}
\label{experimentalcomparing} &&f(t)=\int_0^{\infty}
w(k_2)\frac{k_{1}k_{2}[S]}{2A}\left[e^{(B+A)t}-e^{(B-A)t}\right]dk_2,
\end{eqnarray}
while the weight function $w(k_2)$ is related to the initial
equilibrium distribution as follows,
\begin{eqnarray}
w(k_2)=P^{\rm eq}_{\rm E}\left[x^{-1}(k_2)\right]\frac{dx}{dk_2},
\end{eqnarray}
where $x^{-1}(k_2)$ is the inverse function of $k_2(x)$. In order
to demonstrate the usage of this ``microscopic" interpretation, we
fit our theory by assuming that the potential $V_{\rm E}$ has a
harmonic form with spring constant $k$, \emph{i.e.},
\begin{eqnarray}
\label{Gaussiandistribution}
P_{\rm E}(x)=\frac{1}{\sqrt{2\pi
}\sigma}\exp(-\frac{x^2}{2\sigma^2})
\end{eqnarray}
where $\sigma^2=k_{\rm B}T/k$, and $k_2(x)=a\exp(-bx)$. It might
be the simplest model of Eq.~(\ref{experimentalcomparing}). The
values of the parameters and fitting results are showed in
Fig.~\ref{figure1}. We see that our calculation is satisfactory.
Because Eq.~(\ref{experimentalcomparing}) is almost the same with
previous result~\cite{Kou}, we are not ready to explain the
general behavior of it afresh. In the following part, we will
focus on the general solutions to the coupled diffusion-reaction
equations.

Firstly substituting~\cite{Zusman,Sumi,Zhu}
\begin{eqnarray}
\label{transform} P_{\rm I}(x,t)=g_{\rm I}(x)Q_{\rm I}(x,t)
\end{eqnarray}
into Eq.~(\ref{diffusionreactonequations}), where $g_{\rm I}(x)$,
I= E, ES, and E$^0$ are related to the thermal equilibrium
distributions
\begin{eqnarray}
\label{ted} g_{\rm I}(x)&=&\left[P^{\rm eq}_{\rm
I}(x)\right]^{1/2}\nonumber\\
&=&e^{-\beta V_{\rm I}(x)/2}\left/\left[\int e^{-\beta V_{\rm
I}(x)}dx\right.\right]^{1/2},
\end{eqnarray}
we transform the diffusion reaction equations into an adjoint form
\begin{eqnarray}
\label{adjointformdiffusionreactionequations}
&&\frac{\partial}{\partial
t}Q_{\rm E}(x,t)=-[{\hat H}_{\rm E}+k_{1S}(x)]Q_{\rm E}+k_{-1}'(x)Q_{\rm ES}\nonumber\\
&&\frac{\partial}{\partial
t}Q_{\rm ES}(x,t)=-[{\hat H}_{\rm ES}+k_{3}(x)]Q_{\rm ES}+k_{1S}'(x)Q_{\rm E}\nonumber\\
&&\frac{\partial}{\partial t}Q_{\rm E^0}(x,t)=-{\hat H}_{\rm
E^0}Q_{\rm E^0}+{k_2}'(x)Q_{\rm ES},
\end{eqnarray}
where the new functions $k_{-1}'(x)$, $k_{1S}'(x)$ and $k_2'(x)$
are respectively defined by
\begin{eqnarray}
\label{newreactionrates}
&&k_{-1}'(x)=k_{-1}'g_{\rm ES}/g_{\rm E}(x), \nonumber\\
&&k_{1S}'(x)=k_{1\rm S}g_{\rm E}/g_{\rm ES}(x), \nonumber\\
&&k_2'(x)=k_2g_{\rm ES}(x)/g_{\rm E^0}(x),
\end{eqnarray}
and the Hamiltonian operators are
\begin{eqnarray}
\label{Hamitonianoperators} &&H_{\rm I}=-D_{\rm I}\frac{\partial
^2}{\partial x^2}+\frac{\beta D_{\rm I}}{2}\left[
\frac{\beta}{2}\left(\frac{dV_{\rm
I}}{dx}\right)^2-\frac{d^2V_{\rm I}}{dx^2}\right],
\end{eqnarray}
respectively. We assume that the operators ${\hat H}_{\rm I}$ have
discrete eigenfunctions $|n\rangle_{\rm I}$ (the bound diffusion
assumption), $i.e.$,
\begin{eqnarray}
\label{eigen} {\hat H}_{\rm I}|n\rangle_{\rm I}=\epsilon_{{\rm
I}_n}|n\rangle_{\rm I},\qquad n=0,1,\cdots
\end{eqnarray}
then $g_{\rm I}(x)$ are just the lowest order eigenfunctions
$|0\rangle_{\rm I}$ in the coordinate representation with zero
eigenvalues ($\epsilon_{i_0}=0$). The reader is reminded that the
diffusion information has been included in the eigenvalues, for
instance, given the potentials $V_{\rm I}$ to be harmonic like
Eq.~(\ref{Gaussiandistribution}), then $\epsilon_{{\rm
I}_n}=nk\beta D_{\rm I}$. Defining ${\hat O}_{\rm I}=s+{\hat
H}_{\rm I}+k_i(x)$, here $i=1S$ and 3 respectively correspond to
I=E and ES, and ${\hat O}_{\rm E^0}=s+{\hat H}_{\rm E^0}$, the
Laplace transform solution $Q_{\rm ES}(x,s)$ of
Eq.~(\ref{adjointformdiffusionreactionequations}) with the initial
conditions can be written as
\begin{eqnarray}
\label{LaplacesolutionQES} Q_{\rm ES}(x,s)={\hat O}_{\rm
ES}^{-1}k_{1S}'\frac{1}{{\hat O}_{\rm E}-k_{-1}'{\hat O}_{\rm
ES}^{-1}k_{1S}'}|0\rangle_{\rm E}.
\end{eqnarray}

Although the above calculations are exact formally, we cannot say
more the inverse operator $\hat O_{\rm ES}^{-1}$. Therefore we
employ the decoupled approximation~\cite{Sumi,Zhu}
\begin{eqnarray}
\label{decouplingapproximation} 1\approx {k^j_{{\rm I}_{\rm
eq}}}^{-1}|0\rangle_{\rm I} {}_{\rm I}\langle 0|k_j,
\end{eqnarray}
where $k^j_{{\rm I}_{\rm eq}}={}_{\rm I}\langle 0
|k_j|0\rangle_{\rm I}$. This is would be exact when the
expectation value of the operator
Eq.~(\ref{decouplingapproximation}) is computed in the state
$|0\rangle_{\rm I}$. Using the approximated unit operators in
Eq.~(\ref{waitingtimedistributionLP}) repeatedly, we finally get
the analytical form of $f(s)$ as follows,
\begin{eqnarray}
\label{finalwaitingtimedistributionLP} f(s)=\frac{{}_{\rm
E}\langle |k_3k_{1\rm S}|0\rangle_{\rm E}k^2_{{\rm ES}_{\rm
eq}}/k^3_{{\rm ES}_{\rm eq}}}{s^2\left[1+a^3_{\rm
ES}(s)\right]\left[1+a^{1\rm S}_{\rm E}(s)\right]-{}_{\rm
E}\langle 0|k_3k_{1\rm S}|0\rangle_{\rm E}{}_{\rm ES}\langle
0|k_{-1}k_{1\rm S}|0\rangle_{\rm ES}/k^{1\rm S}_{{\rm E}_{\rm eq}}
k^{3}_{{\rm ES}_{\rm eq}}}
\end{eqnarray}
where
\begin{eqnarray}
\label{asidefinition} a^i_{\rm I}(s)&=&{k^i_{{\rm I}_{\rm
eq}}}^{-1}{}_{\rm I}\langle0|k_i\left(s+
{\hat H}_{\rm I}\right)^{-1}|0\rangle_{\rm I}\\
&=&k^i_{{\rm I}_{\rm eq}}s^{-1}+{k^i_{{\rm I}_{\rm
eq}}}^{-1}\sum\limits_{n=1}^{\infty}(s+\epsilon_{{\rm
I}_n})^{-1}\left|{}_{\rm I}\langle 0|k_i|n\rangle_{\rm I}\right
|^2,\nonumber
\end{eqnarray}
and $i=1S$ and 3 correspond to I=E and ES, respectively. We
immediately see that the waiting time distribution $f(t)$ has a
multiexponential behavior, because the denominator of
Eq.~(\ref{finalwaitingtimedistributionLP}) is a higher order ($\ge
2$) polynomial. For instance, if we truncate the $a^i_{\rm I}(s)$
to $n$th, $f(t)$ should be a sum of 2(n+1) exponential decay
functions. A remarkable conclusion is that, even if the waiting
time distribution function has very complicated multiexponential
decay behavior, the reciprocal of the first moment of
distribution, $\langle t \rangle=-df(s)/ds\left|_{s=0}\right.$
still has a simple MM-like expression,
\begin{eqnarray}
\label{finaMMequation} \frac{1}{\langle t \rangle}=\frac{{\cal K}
[S]}{[S]+{\cal M}},
\end{eqnarray}
where
\begin{eqnarray}
&&{\cal M}=k^3_{{\rm ES}_{\rm eq}}/{\cal F},\nonumber \\&&{\cal
K}=k^3_{{\rm ES}_{\rm eq}}\left(k^3_{{\rm ES}_{\rm eq}}k^1_{{\rm
E}_{\rm eq}}-{}_{\rm ES}\langle0|k_1k_{-1}|0\rangle_{\rm
ES}{}_{\rm E}\langle0|k_1k_3|0\rangle_{\rm E}\left/k^3_{{\rm
ES}_{\rm eq}}k^1_{{\rm E}_{\rm eq}}\right.\right)^2\left/{\cal
F}k^2_{{\rm ES}_{\rm eq}}{}_{\rm E}\langle 0|k_1k_3|0\rangle_{\rm
E}\right.,\nonumber
\end{eqnarray}
and
\begin{eqnarray}
{\cal F}=\left(1+{k^3_{{\rm ES}_{\rm
eq}}}^{-1}\sum\limits_{n=1}^{\infty}\epsilon^{-1}_{{\rm
ES}_n}\left|{}_{\rm ES}\langle 0|k_3|n\rangle_{\rm
ES}\right|^2\right)k^1_{{\rm E_{\rm eq}}}+ {k^1_{{\rm E}_{\rm
eq}}}^{-1}\sum\limits_{n=1}^{\infty}\epsilon^{-1}_{{\rm
E}_n}\left|{}_{\rm E}\langle 0|k_1|n\rangle_{\rm
E}\right|^2k^3_{{\rm ES}_{\rm eq}}.\nonumber
\end{eqnarray}
Here we have separated the substrate concentration $[S]$ from the
rate $k_{1S}(x)$. Under the two limiting cases discussed at the
beginning, Eq.~(\ref{asidefinition}) is approximated to
be~\cite{Zhu}
\begin{eqnarray}
a^i_{I}(s){\rm} \approx
k^i_{{\rm I}_{\rm eq}}s^{-1} \qquad(\textrm{slow reaction limit}),\nonumber \\
a^i_{I}(s){\rm} \approx k_i(x)s^{-1} \qquad(\textrm{nondiffusion
limit}).
\end{eqnarray}
Substituting them into Eq~(\ref{finalwaitingtimedistributionLP})
and making the Laplace transformation, we obtain the same
Eqs~(\ref{slowreactionlimit})
and~(\ref{nondiffusionlimitdistribution}). The general solution
hence well recovers the two limiting cases. Because the decoupling
approximation Eq.~(\ref{decouplingapproximation}) has been proved
to be a good approximation~\cite{Sumi,Zhu}, we conclude that
classic MM equation still is a good approximation even in the
presence of dynamic disorder with arbitrary characteristics.

There are two main contributions in the present work. Firstly we
recover the waiting time distribution $f(t)$ obtained by Xie
\emph{et al.} in quasi-static condition, and  given a microscopic
interpretation of the weight function used by them. But compared
to their complicated algebra calculation and a continuum
approximation involved, our approach is very simple and direct. We
must point out that the current calculations except fitting to the
experiment are independent of specific conformational dynamics.
Second, we get general waiting time distribution
Eq.~(\ref{finalwaitingtimedistributionLP}) with arbitrary dynamic
disorder, and prove that the reciprocal of its first moment still
follows the classic MM equation. Although this conclusion is based
on decoupling approximation, it still is positive because this
approximation has been proved to work well in various systems.
Moreover, it is beyond the quasi-static disorder condition. While
the discrete chemical reaction scheme of Xie \emph{et al} is hard
to achieve because of mathematic difficult. We believe that
Eq.~(\ref{finalwaitingtimedistributionLP}) would be more useful
than Eq.~(\ref{nondiffusionlimitdistribution}) when experiments
are performed on various enzyme molecule under a broad range of
environmental condition.\\
\\This work was supported in part by the National Science Foundation of
China and the National Science Foundation under Grant No.
PHY99-07949.

\begin{figure}[htpb]
\begin{center}
\includegraphics[width=1\columnwidth]{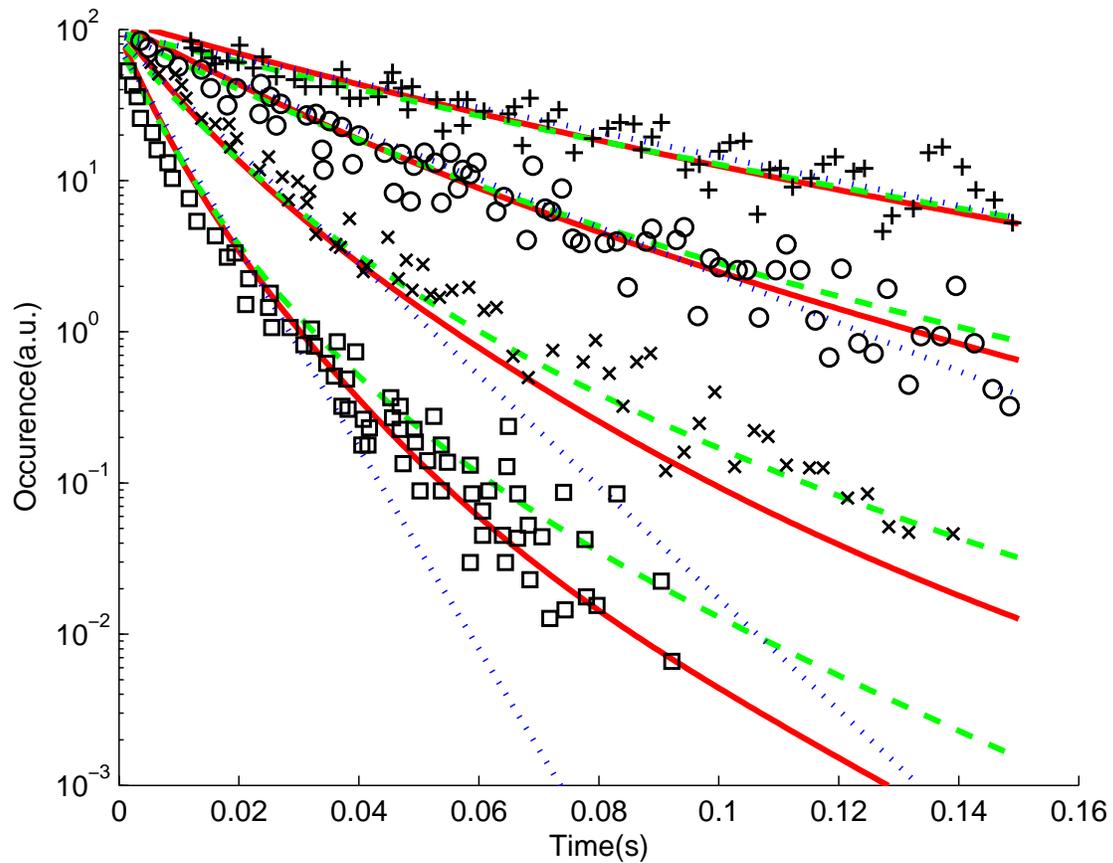}
\label{figure1} \caption{Waiting time distribution vs. substrate
concentration. The dotted and dashed lines and the experiment data
are from Ref.~\cite{English}. The substrate concentrations are 10
$\mu$M (the cross), 20 $\mu$M (the circle), 50 $\mu$M (the time)
and 100 $\mu$M (the square), respectively. The parameters used in
Eq.~(\ref{experimentalcomparing}) are $k_1=5\times
10^7M^{-1}s^{-1}$, $k_{-1}=18300$ $s^{-1}$, $a=904$ $s^{-1}$,
$b=5.0$ nm and $\sigma =0.1$ nm~\cite{liufexp}. } \label{figure1}
\end{center}
\end{figure}


\begin{thebibliography}{99}
\bibitem{liufexp}
The reader is reminded that we do not know the real dimension of
the distance parameters $b$ and $\sigma$ by fitting the existing
experimental data. We appoint them to be \emph{nm} only as a
reference.
\end{thebibliography}
\end{document}